\documentclass{article}
\normalsize
\usepackage{multicol}

\usepackage{cite}
\usepackage{amsfonts}
\usepackage{set space}
\usepackage{graphicx}
\usepackage{mathptmx}  
\usepackage{latexsym}
\usepackage{latexsym}
\usepackage{amsmath}
\usepackage{amsthm}

\title{
Derivation of the cutoff length from the quantum quadratic enhancement of a mass in
vacuum energy constant Lambda
} 
\author{
Kimichika Fukushima
\thanks{E-mail: kimichika1a.fukushima@glb.toshiba.co.jp; km.fukushima@mx2.ttcn.ne.jp 
Phone: +81-90-4602-0490 Phone/Fax: +81-45-831-8881}\\
Advanced Reactor System Engineering Department,\\
Toshiba Nuclear Engineering Services Corporation,\\
8, Shinsugita-cho, Isogo-ku, Yokohama 235-8523, Japan\\
 \\
Hikaru Sato\\
Emeritus, Department of Physics, Hyogo University of Education,\\
Yashiro-cho, Kato-shi, Hyogo 673-1494, Japan
}

\date{             }

\begin{document}

\maketitle

Ultraviolet self-interaction energies in field theory sometimes contain meaningful physical quantities. The self-energies in such as classical electrodynamics are usually subtracted from the rest mass. For the consistent treatment of energies as sources of curvature in the Einstein field equations, this study includes these subtracted self-energies into
vacuum
energy expressed by the constant Lambda (used in such as Lambda-CDM). In this study, the self-energies in electrodynamics and macroscopic classical Einstein field equations are examined, using the formalisms with the ultraviolet cutoff scheme. One of the cutoff formalisms is the field theory in terms of the step-function-type basis functions, developed by the present authors. The other is a continuum theory of a fundamental particle with the same cutoff length. Based on the effectiveness of the continuum theory with the cutoff length shown in the examination, the dominant self-energy is the quadratic term of the Higgs field at a quantum level (classical self-energies are reduced to logarithmic forms by quantum corrections). The cutoff length is then determined to reproduce today's tiny value of Lambda for
vacuum
energy. Additionally, a field with nonperiodic vanishing boundary conditions is treated, showing that the field has no zero-point energy.

\begin{multicols}{2}

\section{Introduction}
\label{sec:Sec1}


Self-interaction energies in field theory, which contain ultraviolet divergences in continuum theory, sometimes reveal meaningful properties in physics
\cite{Lamb,Weiss,Weiss39,Tomo46,Tomo48,Schwin,Feyn,Dys}.
In our previous paper \cite{Fuku84,Fuku14,Fuku16,Fuku17}, we formulated a field theory in terms of the step-function-type basis functions (SFT field theory), which is based on the finite element theory
\cite{Fuku84,Fuku17,FemMW,BenMS}
(the formulation is rather different from that by Bender {\it et al.}), and cuts off high-frequency oscillations of wave functions at short distances. Owing to the space-time continuum and differentiable step-function-type basis functions, this formalism is Poincar\'e covariant and removes ultraviolet divergences at short distances.
The advantage of our formalism is the availability to perform self-energy evaluation.
(We note that the conventional finite element method is widely used \cite{FemMW}.
The validity of theories is of course justified solely by the correctness of the logical deduction.
The support based only on the fact, where an article was published, is insufficient for the true justification of the theory.
The assessment of the theory is beyond the action by authors.)
The meaningful self-energy appears in the Lamb shift \cite{Lamb}, which is caused by finite parts of the self-energy in higher-order terms, and the divergent parts are subtracted from the rest mass.
In contrast, the self-energy also appears in the $\phi^3$ model (the mass is sometimes not renormalized when the mass is a value in vacuum without containing additional interactions).
In our previous paper \cite{Fuku17}, we derived excited states such as meta-stable states at stationary states, which are not always orthogonal to the ground state.
\par
In the Einstein field equations \cite{Eins,LanL}, the rest energy works as a source of the curvature.
The mass renormalization in
such as
electrodynamics subtracts self-energies, which can
be finite using the cutoff scheme. It is then expected that the self-energies are involved in the Einstein field equations.
\par
In our formalism,
four-dimensional space-time is divided into many hyper-octahedrons, whose shape are arbitrary and have the size $\Delta$ (cutoff length) in four-dimensional space-time. For simplicity, we consider three-dimensional space and divide the region into many cubes. The classical wave function
$\phi(x,y,z)$
is expressed in terms of the step-function-type basis functions
$\tilde{\Omega}_p^3(x,y,z)$ in three-dimensional space (the step-function-type basis function in one dimension is defined by Eq. (\ref{eqn:BasD}))
\begin{eqnarray}
\phi(x,y,z)=\sum_{p}\phi_p \tilde{\Omega}_p^3(x,y,z),
\end{eqnarray}
where the basis function takes a value of 1 (one) in a cube (each cubic region is identified by index $p$) and vanishes outside the cube. The coefficient
$\phi_p$
is a constant within the cubic region identified by the index $p$.
\par
Motivated by the above expectation, this paper is aimed at presenting a formulation to include the subtracted self-energies into
vacuum
energy with the constant $\Lambda$ (cosmological constant) \cite{Eins,Fried,Carr92,Carr01,Carr04,Riess,Baker,Perl,Peeb,Padman,Tegm,Lomb,Barr,Rugh,Hobs,Wein,Leut} of the macroscopic classical Einstein field equations.
The self-energy in classical electrodynamics is calculated by the continuum theory with a finite cutoff length.
The self-energy is also derived using the field theory in terms of the step-function-type basis functions, which was developed by the present authors, and the result is compared with that calculated by the continuum theory.
We also examine the curvature (gravitational) self-energy of the fundamental particle with the energy of a rest mass.
Considering the examinations that the self-energies derived in terms of the step-function-type basis functions and that by the continuum theory with the same cutoff length are not so different, the classical self-energies are reduced to the logarithmic forms. However, the self-energy of the Higgs boson has the larger quadratic form.
The derived self-energy is subtracted and involved in the repulsive
vacuum
energy with the constant $\Lambda$.
Under a classical gravitational field, whose strength is small for scales larger than the Planck scale, 
we consider the contribution from the self-energy of a Higgs boson to
vacuum
energy.
The cutoff length is then determined to reproduce the observed
vacuum
energy constant
$\Lambda$.
This theoretical
vacuum
energy constant $\Lambda$ has today's tiny value.
\par
This paper is organized as follows:
Section \ref{sec:Sec2} presents the formalism and analysis procedure.
We exhibit a formalism for the subtraction of the self-energy by including the energy into
vacuum
energy constant
$\Lambda$ (cosmological constant).
Subsequently, the field theory in terms of the step-function-type basis functions is described to derive finite self-energies.
Section \ref{sec:Sec3} examines the self-energy in classical electrodynamics and from the macroscopic classical Einstein field equations. The self-energies are calculated by the continuum theory and the field theory in terms of the step-function-type basis functions.
Section \ref{sec:Sec4} describes the relationship between the subtracted self-energies and
vacuum
energy constant $\Lambda$,
and derive the cutoff length to reproduce
vacuum
energy with the constant $\Lambda$,
followed by Sec. \ref{sec:Sec5}, which summarizes the conclusions.
\par


\section{Formalism for self-energies and the field theory in terms of the step-function-type basis functions}
\label{sec:Sec2}

\subsection{Formalism for the subtraction of the self-energy by involving the energy into
vacuum
energy constant
$\Lambda$}
\label{sec:Sec21}

In this subsection, we present the formalism for the inclusion of subtracted
self-energies
produced by interactions (in such as electrodynamics) into
vacuum
energy constant
$\Lambda$.
Throughout this paper, the notation $x^0=ct$ (c is the velocity of light) is a time coordinate, and $x^i$ are space coordinates, where $x^1=x$, $x^2=y$ and $x^3=z$. The infinitesimal squared distance (according to the notations by
Bjorken and Drell \cite{BJDR}
is denoted as
\begin{eqnarray}
(ds)^2
=g_{\mu \nu} dx^{\mu} dx^{\nu},
\end{eqnarray}
where $g_{\mu \nu}$ is the metric tensor and the indices run over 0, 1, 2 and 3. We use the summation conventions such as
\begin{eqnarray}
g_{\mu \nu}dx^{\nu}=g_{\mu 0} dx^{0}+g_{\mu 1} dx^{1}+g_{\mu 2} dx^{2}+ g_{\mu 3} dx^{3},
\end{eqnarray}
for Greek indices and
\begin{eqnarray}
g_{\mu i}dx^i=g_{\mu 1} dx^1+g_{\mu 2} dx^2+ g_{\mu 3} dx^3,
\end{eqnarray}
for Latin indices.
The metric tensor of $g_{\mu \nu}$ in a flat Minkowski space is given by
\begin{eqnarray}
g_{\mu \nu}=
\left[\begin{array}{cccc}
 1 & 0 & 0 & 0 \\
 0 & -1 & 0 & 0 \\
 0 & 0 & -1 & 0 \\
 0 & 0 & 0 & -1 \\
\end{array}\right].
\end{eqnarray}
The action functional $S_{\rm g}$ for the gravity is expressed by
\begin{eqnarray}
S_{\rm g}=\int {\cal L}_{\rm g} \sqrt{-g} dV_4,
\end{eqnarray}
where
g is the determinant of $g_{\mu \nu}$ denoted as $g=\det(g_{\mu \nu})$,
and $dV_4=dx^0 dx^1 dx^2 dx^3$. For the gravity,
\begin{eqnarray}
{\cal L}_{\rm g}=\frac{-c^3}{16 \pi G} R,
\end{eqnarray}
where $G$ is the gravitational constant and $R$ is the scalar curvature written by
\begin{eqnarray}
R=g^{\mu \nu} R_{\mu \nu},
\end{eqnarray}
with $R_{\mu \nu}$ being defined by
\begin{eqnarray}
R_{\mu \nu}=R^{\rho}_{\mu \rho \nu},
\end{eqnarray}
using the Riemann curvature tensor $R^{\rho}_{\mu \rho \nu}$.
The tensor $R^{\rho}_{\mu \rho \nu}$ is expressed in terms of the Christoffel symbol $\Gamma^{\lambda}_{\mu \nu}$ as
\begin{eqnarray}
R^{\rho}_{\mu \rho \nu}=
\frac{\partial \Gamma^{\rho}_{\nu\sigma}}{\partial x^{\mu}}
-\frac{\partial \Gamma^{\rho}_{\mu\sigma}}{\partial x^{\nu}}
+\Gamma^{\rho}_{\mu \lambda}\Gamma^{\lambda}_{\nu \sigma}
-\Gamma^{\rho}_{\nu \lambda}\Gamma^{\lambda}_{\mu \sigma},
\end{eqnarray}
where
\begin{eqnarray}
\Gamma^{\lambda}_{\mu \nu}
=\frac{1}{2}
g^{\lambda \rho}
(
 \frac{\partial g_{\rho \mu}}{\partial x^{\nu} }
+\frac{\partial g_{\rho \nu}}{\partial x^{\mu} }
+\frac{\partial g_{\mu  \nu}}{\partial x^{\rho}}
).
\end{eqnarray}
\par
Meanwhile, the action functional of the matter is denoted as
\begin{eqnarray}
S_{\rm m}=
\frac{1}{c} \int {\cal L}_{\rm m} \sqrt{-g} dV_4,
\end{eqnarray}
where ${\cal L}_{\rm m}$ is the Lagrangian density of the matter, and the energy-momentum tensor of the matter is obtained from the relation
\begin{eqnarray}
\frac{1}{2}\sqrt{-g}T_{\mu\nu}=
-(
{\partial x_{\rho}}
\frac{\partial\sqrt{-g}{\cal L}_{\rm m}}
{\partial \frac{\partial g^{\mu \nu}}{\partial x^{\rho}}}
-\frac{\partial \sqrt{-g}{\cal L}_{\rm m}}{\partial g^{\mu \nu}}
).
\end{eqnarray}
The variational calculus with respect to $\delta g^{\mu \nu}$ of the total action functional
\begin{eqnarray}
\label{eqn:vari}
\nonumber
\delta S_{\rm g} +\delta S_{\rm m}
\end{eqnarray}
\begin{eqnarray}
\nonumber
=\frac{-c^3}{16\pi G}
\int  (R_{\mu \nu}-\frac{1}{2}g_{\mu\nu}R- \frac{8\pi G}{c^4}T_{\mu \nu})
\delta g^{\mu\nu} \sqrt{-g} dV_4,
\end{eqnarray}
\begin{eqnarray}
\end{eqnarray}
yields the Einstein field equations
\begin{eqnarray}
\label{eqn:Eeq}
R_{\mu\nu}-\frac{1}{2}g_{\mu\nu}R=\frac{8\pi G}{c^4} T_{\mu\nu}.
\end{eqnarray}
\par
The renormalization of the mass by interactions in such as electrodynamics subtracts self-energies from the rest mass. Because the energy of the rest mass produces curvature (gravity), the subtracted energies are included in
vacuum
energy with the constant
$\Lambda$.
In the above equation, we then add the following tensor for the removal of self-energies produced by interactions (in such as electrodynamics):
\begin{eqnarray}
T^{(\rm S)}_{\mu\nu}=\frac{c^4}{8\pi G}g_{\mu\nu}\Lambda^{(\rm S)},
\end{eqnarray}
where $\Lambda^{(\rm S)}$
is regarded as
vacuum
energy constant
$\Lambda$
(cosmological constant).
The Einstein field equations given by Eq. (\ref{eqn:Eeq}) is rewritten as follows
\begin{eqnarray}
\label{eqn:EeqR}
R_{\mu\nu}-\frac{1}{2}g_{\mu\nu}R=\frac{8\pi G}{c^4}
( T_{\mu\nu}-T^{(\rm S)}_{\mu\nu}).
\end{eqnarray}
We then have
\begin{eqnarray}
\label{eqn:EeqL}
R_{\mu\nu}-\frac{1}{2}g_{\mu\nu}R+ g_{\mu\nu} \Lambda^{(\rm S)}=\frac{8\pi G}{c^4}
T_{\mu\nu},
\end{eqnarray}
which corresponds to the Einstein field equations with
vacuum
energy constant
$\Lambda$.
Consequently, subtracted self-energies in interactions are involved in
vacuum
energy with the constant
$\Lambda$.

\subsection{Field theory in terms of the step-function-type basis functions
}
\label{sec:Sec22}

In describing physical quantities at short distances, theories are required to remove ultraviolet divergences. We formulated the field theory \cite{Fuku84,Fuku14,Fuku16,Fuku17}, which is expressed in terms of the step-function-type basis functions to realize the removal of the ultraviolet divergences.
In this subsection, the formalism is described to express fields in terms of the
step-function-type
basis functions in the form used by this paper.
Our described formalism divides four-dimensional real space-time into
hyper-octahedrons
with the arbitrary shapes of boundaries. The hyper-octahedron in real space-time is mapped from a hypercube with flat boundary surfaces in a parameter space-time.
A basis function defined around a center of a hypercube takes a value of unity (one) and vanishes outside the hypercube.
\par
In this paper, the cubic region in three-dimensional space is approximated by the spherical region for simplicity and convenience.
We calculate fields in spherical coordinates and divide the spherical symmetric region into shells.
The results can be generalized to the case in which the region is divided into many hyper-octahedrons with arbitrary shapes.
Grid (lattice) points along the radial $r$-axis
($r=(x^2+y^2+z^2)^{1/2}$)
are denoted as
$r_1$, $r_2$,…, $r_k$,…, $r_{N_{r}+1}$,
with $k=1, 2,$…, $N_{r}+1$,
where $N_{r}$ is the number of lattice points
and $k=N_{r}+1$ is the lattice index for a boundary.
We here set the radial cutoff length
$\Delta_{\rm h}$ 
(corresponding to the cutoff length $\Delta$ with $\Delta=2\Delta_{\rm h}$)
to the lattice spacing as
$\Delta_{\rm h}$
$=r_{k}-r_{k-1}$
and define the notations
$r_{k-1/2}=r_k$
$-\Delta_{\rm h}/2$
and
$r_{k+1/2}=r_k$
$+\Delta_{\rm h}/2$.
The step-function-type basis function used is defined by
\begin{eqnarray}
\label{eqn:BasD}
{\tilde{\Omega}}^{E}_{k}(r)
=
\left\{\begin{array}{ll}
1 & \mbox{ for $r_{k-1/2} \leq r < r_{k+1/2}$}\\
\mbox{ } & \mbox{ } \\
0 & \mbox{ for $r < r_{k-1/2}$ or
$r \geq r_{k+1/2}$}
\end{array}\right. ,
\end{eqnarray}
which has the properties that
\begin{eqnarray}
\label{eqn:Basd1}
\frac{d {\tilde{\Omega}}^{E}_{k}(r)}{d r}|_{r=r_{k-1/2}}
=\delta(r-r_{k-1/2}),
\end{eqnarray}
\begin{eqnarray}
\label{eqn:Basd2}
\frac{d {\tilde{\Omega}}^{E}_{k}(r)}{d r}|_{r=r_{k+1/2}}
=-\delta(r-r_{k+1/2}),
\end{eqnarray}
where $\delta(r)$ is the Dirac delta function.
\par
The field $\phi(r)$ in spherical coordinates is transformed to
\begin{eqnarray}
\label{eqn:tranu}
u(r)=r \phi(r),
\end{eqnarray}
and this wave function $u(r)$ is expressed in terms of basis functions defined by Eq. (\ref{eqn:BasD}) as
\begin{eqnarray}
\label{eqn:BasE}
u(r)=\sum_k u_k {\tilde{\Omega}}^{E}_k (r).
\end{eqnarray}
Thus, we have prepared the formalism to analyze self-energies in the next section.

\section{
Analysis of self-energies from interactions by classical fields
}
\label{sec:Sec3}

\subsection{
Self-energy and mass renormalization in classical electrodynamics by the continuum theory}
\label{sec:Sec31}

This subsection examines and summarizes the self-energy in classical electrodynamic interactions using the continuum theory
\cite{LanL,Rohr62,Dira38,Rohr60,Rohr61}.
The mass density $\mu_{\rm m}$ of a fundamental particle with a mass $m_{\rm E}$ and size $R_{\rm E}$ is denoted as
\begin{eqnarray}
\label{eqn:DmaMQss}
\mu_{\rm m}=\frac{m_{\rm E}}{
(4\pi/3)
({R_{\rm E}})^3}.
\end{eqnarray}
We divide three-dimensional (3D) space into identical cubic elements, 
which were considered in Subsec. \ref{sec:Sec22}. The cubic region is approximated by a spherical region with radius $R_{0}$.
The charge $Q$ and mass $M$ of the spherical region occupied by the fundamental particle are expressed by
\begin{eqnarray}
\label{eqn:DchMQge}
M=(4\pi/3) (R_{0})^3\mu_m,
\hspace{4ex}
Q=(4\pi/3) (R_{0})^3\rho^{\rm (e)},
\end{eqnarray}
respectively, where $\rho^{\rm (e)}$ is the charge density.
A radial cutoff length
$\Delta_{\rm h}=\Delta/2=R_{0}$ in spherical coordinates, corresponding to the cutoff length $\Delta$,
is introduced for simplicity and convenience.
From the conventional energy-momentum tensor of electrodynamics, the self-energy of the static electric field has the form
\begin{eqnarray}
\label{eqn:CESE}
E^{\rm (e)}_{\rm C}= \int dV_{3}
(
\frac{1}{2}
)|{\bf E}|^2,
\end{eqnarray}
where
$dV_{3}=dxdydz$.
The classical electric field
${\bf E}$ 
is produced as 
${\rm div} ({\bf E})= \rho^{\rm (e)}$
from the electric charge density
$\rho^{\rm (e)}$
and is written by
${\bf E}=-\nabla \phi^{\rm (e)}$,
where $\phi^{({\rm e})}$ is the electric potential and satisfies
\begin{eqnarray}
\label{eqn:QEQ}
-
\nabla^2 \phi^{({\rm e})}=
\rho^{({\rm e})}.
\end{eqnarray}
The above self-energy
\begin{eqnarray}
E^{(\rm e)}_{\rm C}=- \int dV_{3} (
\frac{1}{2}
){\bf E}\cdot \nabla \phi^{\rm (e)},
\end{eqnarray}
becomes, using Gauss's theorem and integration by parts,
\begin{eqnarray}
\label{eqn:CE2SE}
E^{\rm (e)}_{\rm C}= \int dV_{3} (\frac{1}{2}) \rho^{({\rm e})} \phi^{\rm (e)}.
\end{eqnarray}
\par
We consider the case, in which the charges exist in the region $r \leq R_{\rm 0}$ and $\rho^{({\rm e})}=0$ for $r > R_{\rm 0}$ using $Q$ in Eq. (\ref{eqn:DchMQge}). Gauss's theorem for Eq. (\ref{eqn:QEQ})
then gives the following potential
\begin{eqnarray}
\label{eqn:poEto}
\phi^{\rm (e)}(r)=\frac{Q}{
4 \pi
r
}
\hspace{3ex}\mbox{ for } r>R_{\rm 0}.
\end{eqnarray}
\par
Similarly 
for $r \leq R_{\rm 0}$, we have
\begin{eqnarray}
4 \pi r^2 (-\frac{\phi^{\rm (e)}(r)}{dr})=
\frac{4 \pi r^3}{3}
\rho^{\rm (e)}
\end{eqnarray}
yielding
\begin{eqnarray}
-\frac{\phi^{\rm (e)}(r)}{dr}=
\frac{1}{3} 
r
\rho^{({\rm e})}
\end{eqnarray}
followed by
\begin{eqnarray}
\label{eqn:poEti}
\nonumber
\phi^{\rm (e)}(r)=\int  dr^{\prime} (-\frac{\phi^{\rm (e)}(r^\prime)}{dr^{\prime}}) =
\frac{1}{6}
r^2
\rho^{({\rm e})}
\hspace{3ex}\mbox{ for } r \leq R_{\rm 0}.
\end{eqnarray}
\begin{eqnarray}
\end{eqnarray}
To connect $\phi^{\rm (e)}(r)$ for $r \leq R_{\rm 0}$ in Eq. (\ref{eqn:poEti}) continuously with that in Eq. (\ref{eqn:poEto}) for $r > R_{\rm 0}$ at $r=R_{\rm 0}$, we shift $\phi^{\rm (e)}(r)$ in Eq. (\ref{eqn:poEti}) to
\begin{eqnarray}
\label{eqn:poEti2}
\nonumber
\phi^{\rm (e)}(r)=
\frac{1}{6}
r^2
\rho^{({\rm e})}
-
\frac{1}{6}
R_{\rm 0}^2
\rho^{({\rm e})}
+
\frac{Q}
{
4 \pi
R_{\rm 0}
}
\hspace{3ex}\mbox{ for } r \leq R_{\rm 0}.
\end{eqnarray}
\begin{eqnarray}
\end{eqnarray}
\par
Using Eqs. (\ref{eqn:CE2SE}), (\ref{eqn:poEti2}) and $Q$ in Eq. (\ref{eqn:DchMQge}),
we obtain the self-energy by the classical electric interaction in the continuum theory
\begin{eqnarray}
\label{eqn:CSE}
\nonumber
E^{(\rm e)}_{\rm C}
=\frac{1}{2}\int_0^{R_{\rm 0}} dr (4 \pi) r^2 \rho^{\rm (e)} \phi^{\rm(e)} (r)
\end{eqnarray}
\begin{eqnarray}
\nonumber
=\frac{1}{2}(
\frac{4 \pi}{6}\rho^{\rm (e)}
\frac{R_{\rm 0}^5}{5}
\rho^{({\rm e})}
-
\frac{1}{6}
R_{\rm 0}^2
Q \rho^{\rm (e)}
+
\frac{QQ}{4 \pi R_{\rm 0}}
)
\end{eqnarray}
\begin{eqnarray}
\nonumber
=\frac{1}{2}(
 \frac{3}{10}
\frac{QQ}{4 \pi R_{\rm 0}}
-\frac{1}{2}
\frac{QQ}{4 \pi R_{\rm 0}}
+
\frac{QQ}{4 \pi R_{\rm 0}}
)
\end{eqnarray}
\begin{eqnarray}
=\frac{1}{2}
(\frac{4}{5} \frac{QQ}{
4 \pi
R_{\rm 0}
}).
\end{eqnarray}
\par
Under an external force
${\bf f}_{\rm e}$,
the classical Newtonian equation of motion for the above charged object with a small velocity ${\bf v_{\rm C}}$ compared to the speed of light $c$ is expressed by (small magnetic contributions are dropped)
\cite{Rohr62,Dira38,Rohr60,Rohr61}
\begin{eqnarray}
\label{eqn:EQM}
M
\frac{d {\bf v}_{\rm C}}{d t}={\bf f}_{\rm e} + \int dV_3 (\rho^{\rm (e)}{\bf E}),
\end{eqnarray}
where $M$
is the mass of the charged object
in Eq. (\ref{eqn:DmaMQss}).
Using the self-energy in Eq. (\ref{eqn:CE2SE}),
the lower-order terms expanded with respect to $1/c$ amounts to
\begin{eqnarray}
\label{eqn:EMLO}
M
\frac{d {\bf v}_{\rm C}}{d t}=
{\bf f}_{\rm e}
-\frac{4}{3 c^2} E^{\rm (e)}_{\rm C}
\frac{d {\bf v}_{\rm C}}{d t},
\end{eqnarray}
which results in
\begin{eqnarray}
\label{eqn:EMRL}
(M
+\frac{4}{3 c^2} E^{\rm (e)}_{\rm C}
)
\frac{d {\bf v}_{\rm C}}{d t}=
{\bf f}_{\rm e}.
\end{eqnarray}
(The relativistic version was given by Dirac and Rohrlich, where the factor 1 (one) appears corresponding to the above factor 4/3
\cite{Rohr62,Dira38,Rohr60,Rohr61}.)
Due to the requirement from the continuum relativistic theory, the fundamental particle is considered to be pointlike.
Then, the above self-energy diverges, which is why mass renormalization is required in electrodynamics.
In mass renormalization, the self-energy is subtracted from the term with the mass.
\par

\subsection{Self-energy derivation for classical electrodynamics using the field theory in terms of the step-function-type basis functions \label{sec:Sec32}}

In contrast to the analysis of Subsec. \ref{sec:Sec31}, this subsection studies the self-energy of the same object in Subsec. \ref{sec:Sec31} in classical electrodynamic equations, using the step-function-type basis functions.
As mentioned in Subsec. \ref{sec:Sec22} and by Eqs. (\ref{eqn:DmaMQss})-(\ref{eqn:DchMQge}), we divide three-dimensional space into cubic elements with the cutoff length $\Delta$, and each cubic region is approximated by a sphere.
The action functional for the electric field $\phi^{\rm (e)}(r)$
can be written in the form
(considering the form 
$-\nabla^2 \phi^{({\rm e})}-\rho^{({\rm e})}=0$
on the left in Eq. (\ref{eqn:QEQ}))
\begin{eqnarray}
\label{eqn:AcNgr}
\nonumber
S_{\rm f}^{\rm(e)}=
-
\frac{1}{2}\int
dxdydz
\end{eqnarray}
\begin{eqnarray}
\nonumber
\times
[\phi^{\rm (e)}(
x,y,z
)
(
 \frac{\partial ^2}{\partial^2 x}
+\frac{\partial ^2}{\partial^2 y}
+\frac{\partial ^2}{\partial^2 z}
) \phi^{\rm (e)}
(
x,y,z
)]
\end{eqnarray}
\begin{eqnarray}
=
-
\frac{1}{2} \int dr (4 \pi r^2) \phi^{\rm (e)}(r)(
 \frac{d ^2}{d^2 r}
 +\frac{2}{r}\frac{d}{d r}
 )\phi^{\rm (e)}(r).
\end{eqnarray}
Using the transformed potential
\begin{eqnarray}
\label{eqn:rQu}
u^{\rm (e)}(r)=r\phi^{\rm (e)}(r),
\end{eqnarray}
as Eq. (\ref{eqn:tranu})
for spherical coordinates and integration by parts, we find
\begin{eqnarray}
\label{eqn:AcQrr}
\nonumber
S_{\rm f}^{\rm (e)}
=
-
\frac{1}{2} \int dr (4 \pi r^2) \frac{u^{\rm (e)}}{r}
\frac{1}{r}
[ \frac{d ^2}{d^2 r}
 u^{\rm (e)}(r)]
\end{eqnarray}
\begin{eqnarray}
\nonumber
=\frac{-1}{2} \int dr (4 \pi)
[\frac{d ^2}{d^2 r}u^{\rm (e)}(r)]
\end{eqnarray}
\begin{eqnarray}
=\frac{
1
}{2} \int dr (4 \pi)
\frac{du^{\rm (e)}(r)}{dr}
\frac{du^{\rm (e)}(r)}{dr}.
\end{eqnarray}
\par
As in Subsec. \ref{sec:Sec22} and by Eqs. (\ref{eqn:DmaMQss})-(\ref{eqn:DchMQge}), we divide three-dimensional space, containing the above sphere with radius $R_{0}$ centered at the origin in spherical coordinates, into shells (the number of cells enclosing the central sphere is $N_r-1$). The radial width (lattice spacing implying the radial cutoff length) of each shell is $\Delta_{\rm h}$, which is equal to the radius $R_{0}$ of the enclosed central sphere.
As Eq. (\ref{eqn:BasE}), the above wave function $u^{\rm (e)}(r)$ is expressed by
\begin{eqnarray}
\label{eqn:EQp}
u^{\rm (e)}(r)=\sum_k u^{\rm (e)}_k {\tilde{\Omega}}^{E}_k (r),
\end{eqnarray}
in terms of the step-function-type basis functions ${E}_k (r)$ in Eq. (\ref{eqn:BasD}).
From Eq. (\ref{eqn:AcQrr}), it follows that ($k, K=$1, 2,…, $N_{{r}+1}$)
\begin{eqnarray}
\nonumber
S_{\rm f}^{\rm (e)}
=\sum_{k,K} 
[\frac{
1
}{2} \int dr (4 \pi) 
u^{\rm (e)}_k u^{\rm (e)}_K
(\frac{\partial \tilde{\Omega}^{E}_k (r)}{d r}
\frac{\partial \tilde{\Omega}^{E}_K (r)}{d r})
].
\end{eqnarray}
\begin{eqnarray}
\end{eqnarray}
Using Eqs. (\ref{eqn:Basd1}) and
(\ref{eqn:Basd2}), 
$S_{\rm f}^{\rm (e)}$ above is decomposed into
\begin{eqnarray}
\label{eqn:sgQt}
S_{\rm f}^{\rm (e)}=
S_{\rm f}^{\rm (e)--}
+S_{\rm f}^{\rm (e)-+}
+S_{\rm f}^{\rm (e)+-}
+S_{\rm f}^{\rm (e)
++
},
\end{eqnarray}
where
\begin{eqnarray}
\label{eqn:sgQnn}
\nonumber
S_{\rm f}^{\rm (e)--}=
\end{eqnarray}
\begin{eqnarray}
\nonumber
=4
\pi\frac{1}{2}
\sum_{k,K} u^{\rm (e)}_k u^{\rm (e)}_K \int dr
[\delta(r-r_{k-1/2})\delta(r-r_{K-1/2})],
\end{eqnarray}
\begin{eqnarray}
\end{eqnarray}
\begin{eqnarray}
\nonumber
S_{\rm f}^{\rm (e)-+}=
\end{eqnarray}
\begin{eqnarray}
\label{eqn:sgQnp}
\nonumber
=4
\pi \frac{1}{2}
\sum_{k,K} u^{\rm (e)}_k u^{\rm (e)}_K \int dr
[\delta(r-r_{k-1/2})\delta(r-r_{K+1/2})],
\end{eqnarray}
\begin{eqnarray}
\end{eqnarray}
\begin{eqnarray}
\nonumber
S_{\rm f}^{\rm (e)+-}=
\end{eqnarray}
\begin{eqnarray}
\label{eqn:sgQpn}
\nonumber
=4
\pi \frac{1}{2}
\sum_{k,K}u^{\rm (e)}_k u^{\rm (e)}_K \int dr
[\delta(r-r_{k+1/2})\delta(r-r_{K-1/2})],
\end{eqnarray}
\begin{eqnarray}
\end{eqnarray}
\begin{eqnarray}
\nonumber
S_{\rm f}^{\rm (e)++}=
\end{eqnarray}
\begin{eqnarray}
\label{eqn:sgQpp}
\nonumber
=4
\pi \frac{1}{2}
\sum_{k,K} u^{\rm (e)}_k u^{\rm (e)}_K \int dr
[\delta(r-r_{k+1/2})\delta(r-r_{K+1/2})].
\end{eqnarray}
\begin{eqnarray}
\end{eqnarray}
\par
With the help of the lattice spacing $\Delta_{\rm h}$ mentioned above Eq. (\ref{eqn:BasD}), the element such as $S_{\rm f}^{\rm (e)--}$ in Eq. (\ref{eqn:sgQnn}) is reduced to
\begin{eqnarray}
\label{eqn:sgQQnp}
\nonumber
S_{\rm f}^{\rm (e)--}
=4
\pi \frac{1}{2}
\sum_{k,K} u^{\rm (e)}_k u^{\rm (e)}_K
[\delta(r_{k-1/2}-r_{K-1/2})]
\end{eqnarray}
\begin{eqnarray}
\nonumber
=4
\pi \frac{1}{2} 
\sum_{k,K} 
\frac{\Delta_{\rm h}}{\Delta_{\rm h}}
u^{\rm (e)}_k u^{\rm (e)}_K 
[\delta(r_{k-1/2}-r_{K-1/2})]
\end{eqnarray}
\begin{eqnarray}
\nonumber
=4
\pi \frac{1}{2}
\end{eqnarray}
\begin{eqnarray}
\nonumber
\times
\sum_{k}
\frac{1}{\Delta_{\rm h}}\int d r_{K-1/2}
\{ u^{\rm (e)}_k u^{\rm (e)}_K 
[\delta(r_{k-1/2}-r_{K-1/2})] \}
\end{eqnarray}
\begin{eqnarray}
=4
\pi \frac{1}{2} \frac{1}{\Delta_{\rm h}} \sum_{k,K}
u^{\rm (e)}_k u^{\rm (e)}_K \delta_{k,K},
\end{eqnarray}
where $\delta_{k,K}$ is the Kronecker delta. By similar calculations for the elements of $S_{\rm f}^{\rm (\rm e)}$ given by Eqs. (\ref{eqn:sgQnp})-(\ref{eqn:sgQpp}), the total $S_{\rm f}^{\rm (e)}$ in Eq. (\ref{eqn:sgQt}) amounts to
\begin{eqnarray}
\nonumber
\label{eqn:DEQg}
S_{\rm f}^{\rm (e)}
=\frac{4\pi}{2}
\frac{1}{\Delta_{\rm h}}
\sum_{k,K}
(-u^{\rm (e)}_k u^{\rm (e)}_{K-1} \delta_{k,K-1}
\end{eqnarray}
\begin{eqnarray}
+2u^{\rm (e)}_k u^{\rm (e)}_K \delta_{k,K}-u^{\rm (e)}_k u^{\rm (e)}_{K+1} \delta_{k,K+1}).
\end{eqnarray}
\par
On the other hand, the action functional of the electric charge of the matter for spherical coordinates is expressed using $u^{\rm (e)}(r)$ in Eq. (\ref{eqn:rQu}) by 
(considering also the form 
$-\nabla^2 \phi^{({\rm e})}-\rho^{({\rm e})}=0$
on the left in Eq. (\ref{eqn:QEQ}))
\begin{eqnarray}
\nonumber
S_{\rm m}^{\rm (e)}=
-
\int dr (4 \pi) r^2
\rho^{({\rm e})}
\phi^{\rm (e)}(r)
\end{eqnarray}
\begin{eqnarray}
=
-
\int dr (4 \pi) r
\rho^{({\rm e})}
u^{\rm (e)}(r).
\end{eqnarray}
Subsequently, 
by the expansion of $u^{\rm (e)}(r)$ given by Eq.
(\ref{eqn:rQu}) 
in terms of basis functions denoted in Eq. (\ref{eqn:BasD}), the above action becomes
\begin{eqnarray}
\label{eqn:SmQatb}
\nonumber
S_{\rm m}^{\rm (e)}=
-
\int dr (4 \pi) r
\rho^{({\rm e})}
\sum_k u^{\rm (e)}_k {\tilde\Omega}_k (r)
\end{eqnarray}
\begin{eqnarray}
=
-
(4 \pi)
\rho^{({\rm e})}
\sum_k u^{\rm (e)}_k \frac{r_{k+1/2}^2-r_{k-1/2}^2}{2}.
\end{eqnarray}
Because $R_{0}=\Delta_{\rm h}$ as mentioned below Eq. (\ref{eqn:DmaMQss}),
\begin{eqnarray}
R_{0}=r_{k+1/2}|_{k=1}=r_{1+1/2},
\end{eqnarray}
which implies $\rho^{\rm (e)}=0$ for $k>1$ (the index 1 is one) in Eq. (\ref{eqn:SmQatb}),
and $r_{k-1/2}=0$ (or $r_{k-1/2}=\epsilon$ with $\epsilon \rightarrow 0$ after the calculation).
Using $Q$ in Eq. (\ref{eqn:DchMQge}) and $R_{0}=\Delta_{\rm h}$, we have
\begin{eqnarray}
\label{eqn:DEQm}
\nonumber
S_{\rm m}^{\rm (e)}
=
-
\rho^{({\rm e})}
\sum_k u^{\rm (e)}_k \frac{3}{2}\frac{4 \pi}{3}
\frac{(\Delta_{\rm h})^3}{\Delta_{\rm h}}
\delta_{k,1}
\end{eqnarray}
\begin{eqnarray}
=
-
\sum_k u^{\rm (e)}_k
\frac{3}{2}
\frac{1}{\Delta_{\rm h}} 
Q
\delta_{k,1},
\end{eqnarray}
where $\delta_{k,1}$ is the Kronecker delta (the index 1 is one).
\par
From Eqs. (\ref{eqn:DEQg}) and (\ref{eqn:DEQm}), the variational calculus with respect to $u^{\rm (e)}_k$
\begin{eqnarray}
\delta S_{\rm f}^{\rm (e)} + \delta S_{\rm m}^{\rm (e)}=0,
\end{eqnarray}
leads to
\begin{eqnarray}
\label{eqn:QPrim}
\nonumber
-
\frac{1}{\Delta_{\rm h}}
(u^{\rm (e)}_{k-1}-2u^{\rm (e)}_k+u^{\rm (e)}_{k+1})
=\frac{3}{2}
\frac{Q
^{\prime}
}
{\Delta_{\rm h}} \delta_{k,1}
\end{eqnarray}
\begin{eqnarray}
\hspace{30ex}
\mbox{ with } Q^{\prime}=\frac{Q}{4 \pi}.
\end{eqnarray}
This equation is equivalent to
\begin{eqnarray}
\label{eqn:DeQN}
\frac{
u^{\rm (e)}_{k-1}-2u^{\rm (e)}_k+u^{\rm (e)}_{k+1}}
{(\Delta_{\rm h})^2}
=-
\frac{3}{2}
\frac{Q
^{\prime}
}
{(\Delta_{\rm h})^2} \delta_{k,1} ,
\end{eqnarray}
corresponding to Eq. (\ref{eqn:QEQ}) for $\phi^{\rm (e)} (r)=ru^{\rm (e)}(r)$.
\par
We then have
\begin{eqnarray}
\label{eqn:DeQQ0}
u^{\rm (e)}_{k-1}- 2u^{\rm (e)}_k +u^{\rm (e)}_{k+1}=0
\hspace{4ex} \mbox{ for } k>1,
\end{eqnarray}
which is rewritten by
\begin{eqnarray}
\label{eqn:DeQN00}
u^{\rm (e)}_{k-1}-u^{\rm (e)}_k=u^{\rm (e)}_k-u^{\rm (e)}_{k+1}
\hspace{4ex} \mbox{ for } k>1.
\end{eqnarray}
Additionally, for the boundary $r_{k-1/2}$ with $k=1$ (the index is one)
\begin{eqnarray}
r_{1-1/2}=\epsilon>0,
\end{eqnarray}
(we set $\epsilon \rightarrow 0$ after the calculation), the basis function is not given in the region for $r<0$. Considering this boundary for Eq. (\ref{eqn:DeQN}), we obtain
\begin{eqnarray}
\label{eqn:DeQIB0}
\frac{
-2u^{\rm (e)}_k+u^{\rm (e)}_{k+1}}
{(\Delta_{\rm h})^2}
=-
\frac{3}{2}\frac{Q
^{\prime}
}{(\Delta_{\rm h})^2}
\hspace{4ex} \mbox{ for } k=1.
\end{eqnarray}
In contrast, using $Q
^{\prime}
$ in Eq. (\ref{eqn:DchMQge}) and $R_{0}=\Delta_{\rm h}$
for the charge, the outer boundary condition imposed is
\begin{eqnarray}
\label{eqn:OBQD}
u^{\rm (e)}_{N+1}=Q
^{\prime},
\end{eqnarray}
which implies $\phi^{\rm (e)}(r_{N+1})=Q
^{\prime}
/r_{N+1}$ in Eq. (\ref{eqn:rQu}). Then, Eq. (\ref{eqn:DeQN00}) becomes
\begin{eqnarray}
\label{eqn:DeQOBN}
u^{\rm (e)}_{N-1}-u^{\rm (e)}_N=u^{\rm (e)}_{N}-Q
^{\prime}.
\end{eqnarray}
\par
We consider a solution that takes
\begin{eqnarray}
\label{eqn:DeQOB3}
u^{\rm (e)}_{N}=(\beta-1)Q
^{\prime}
+Q
^{\prime}
\hspace{4ex} \mbox{ for } k = N,
\end{eqnarray}
where $\beta$ is a constant to be determined below.
Equations
(\ref{eqn:DeQN00}) and
(\ref{eqn:OBQD})-(\ref{eqn:DeQOB3})
provide
\begin{eqnarray}
\label{eqn:DeQOB4}
\nonumber
u^{\rm (e)}_{N-1}=[(\beta-1)Q
^{\prime}
+Q
^{\prime}
]
\end{eqnarray}
\begin{eqnarray}
\nonumber
+[(\beta-1)Q
^{\prime}
+Q
^{\prime}
-Q
^{\prime}
\end{eqnarray}
\begin{eqnarray}
=2(\beta-1)Q
^{\prime}
+Q
^{\prime},
\end{eqnarray}
\begin{eqnarray}
\nonumber
\label{eqn:DeQOB5}
u^{\rm (e)}_{N-2}=[2(\beta-1)Q
^{\prime}
+Q
^{\prime}
]
\end{eqnarray}
\begin{eqnarray}
\nonumber
+[2(\beta-1)Q
^{\prime}
+Q
^{\prime}
-[(\beta-1)Q
^{\prime}
+Q
^{\prime}
]
\end{eqnarray}
\begin{eqnarray}
=3(\beta-1)Q
^{\prime}
+Q
^{\prime}.
\end{eqnarray}
Using Eqs. (\ref{eqn:DeQQ0}),
the sequential manipulation results in
\begin{eqnarray}
\label{eqn:DeQOB6}
u^{\rm (e)}_{k}=[(N-k
+1
)(\beta-1)]Q
^{\prime}
+Q
^{\prime}
\hspace{4ex} \mbox{ for } k > 1.
\end{eqnarray}
Because the above solution diverges unless
$\beta=1$
for $k=2$, we derive the following solution, by setting $\beta=1$
and using
$Q^{\prime}=Q/(4 \pi)$ 
in Eq. (\ref{eqn:QPrim}),
\begin{eqnarray}
\label{eqn:DeQOB7}
u^{\rm (e)}_{k}=Q^{\prime}
=\frac{Q}{4 \pi}
\hspace{4ex} \mbox{ for } k > 1.
\end{eqnarray}
Furthermore, from Eqs.
(\ref{eqn:DeQIB0}) 
and (\ref{eqn:DeQOB7})
as well as 
$Q^{\prime}=Q/(4 \pi)$
in Eq. (\ref{eqn:QPrim}),
we have the solution
(at the remaining point)
for $k=1$
\begin{eqnarray}
\label{eqn:DeQNIB}
u^{\rm (e)}_1
=\frac{1}{2}(
\frac{3}{2}Q
^{\prime}
+Q
^{\prime}
)
=\frac{5}{4}Q
^{\prime}
=\frac{5}{4}\frac{Q}
{4 \pi}
\hspace{4ex} \mbox{ for } k=1.
\end{eqnarray}
\par
Thus, from Eqs. (\ref{eqn:CE2SE}), (\ref{eqn:rQu}) and (\ref{eqn:DeQNIB}) with
$2r_1=\Delta_{\rm h}=R_{0}=\Delta /2$,
we obtain the following classical electric self-energy in the region with the cutoff length $\Delta$ (whose volume
$\Delta ^3$ is approximated by
$
(4\pi/3)(\Delta_{\rm h})^3
$
with the charge density $\rho^{\rm (e)}$ and charge $Q$ in Eq. (\ref{eqn:DchMQge})
for this volume)
\begin{eqnarray}
\label{eqn:EQOmega}
\nonumber
E^{\rm (e)}_{\Omega}
=
\frac{1}{2}
\frac{4 \pi}{3}
(\Delta_{\rm h})^3
\rho^{\rm (e)}
\frac{5}{4}Q
(\frac{1}
{
4 \pi
r_1
}
)
\end{eqnarray}
\begin{eqnarray}
\nonumber
=
\frac{1}{2}
\frac{5}{4}
\frac{QQ}{
4 \pi
r_1}
=\frac{1}{2}
[
\frac{5}{2} 
\frac{QQ}
{
(4 \pi)
2r_1
}
]
\end{eqnarray}
\begin{eqnarray}
=\frac{1}{2}(
\frac{5}{2} 
\frac{QQ}{
4 \pi
\Delta_{\rm h}})
=\frac{1}{2}
[
\frac{5}{2} 
\frac{QQ}{
(4 \pi)
(\Delta/2)}
].
\end{eqnarray}

\subsection{
Self-energy in macroscopic classical Einstein field equations}
\label{sec:Sec33}

This subsection presents the analysis of the curvature self-energy in the Einstein field equations.
Although the gravitational field is different from the charged particle fields, we treat the Newtonian approximation case, which is similar to the charged particle case.
When the renormalization is difficult in this case, it is possible to use the cutoff length.
The self-energy is first evaluated by the continuum field theory.
Subsequently,
the self-energy is evaluated using the formalism in terms of the step-function-type basis functions.
As described by Landau and Lifshitz \cite{LanL} (owing to the negligible contributions of higher-order terms with respect to $1/c$ in the Lagrangian with $c$ being the velocity of light), the Newtonian approximation, within the scheme of the Einstein field equations for the matter with the slow velocities compared to $c$, has the infinitesimal squared distance expressed by
\begin{eqnarray}
\nonumber
\label{eqn:New0}
(ds)^2=g_{\mu\nu}dx^{\mu}dx^{\nu}=(\eta_{\mu\nu}+h_{\mu\nu})dx^{\mu}dx^{\nu}
\end{eqnarray}
\begin{eqnarray}
\nonumber
=(1+2\frac{\phi}{c^2})(dx^0)^2
\end{eqnarray}
\begin{eqnarray}
\hspace{11ex}
-(1-2\frac{\phi}{c^2})[(dx^{1})^2+(dx^{2})^2+(dx^{3})^2],
\end{eqnarray}
where
\begin{eqnarray}
\nonumber
\eta_{\mu \nu}=
\left[\begin{array}{cccc}
 1 & 0 & 0 & 0 \\
 0 & -1 & 0 & 0 \\
 0 & 0 & -1 & 0 \\
 0 & 0 & 0 & -1 \\
\end{array}\right],
\hspace{5ex}
|h_{\mu\nu}| \ll 1,
\end{eqnarray}
\begin{eqnarray}
\label{eqn:New1}
g_{00}=
1+2\frac{\phi}{c^2},
\end{eqnarray}
and $\phi$ is the Newtonian potential. We note that the Newtonian potential (field) $\phi$ is distinguished from electric field $\phi^{\rm (e)}$.
Letting $\mu_{\rm m}$ be the density of the mass, we have the energy-momentum tensor
\begin{eqnarray}
\label{eqn:New2}
T_{\mu}^{\nu}
=
\left\{\begin{array}{ll}
\mu_{\rm m}
c^2 & \mbox{ for $\mu=0$ and $\nu=0$}\\
\mbox{ } & \mbox{  }                      \\
0
& \mbox{ for $\mu \neq 0$ or $\nu \neq 0$ }
\end{array}\right. .
\end{eqnarray}
\par
It is known that the field equation Eq. (\ref{eqn:Eeq}) is rewritten by
\begin{eqnarray}
\label{eqn:Eeqrw}
R_{\mu}^{\nu}
=\frac{8 \pi G}{c^4}(T_{\mu}^{\nu}-\frac{1}{2}\delta_{\mu}^{\nu}T),
\end{eqnarray}
where $\delta_{\mu}^{\nu}$ is unit tensor and
\begin{eqnarray}
\label{eqn:New3}
T=g^{\mu \nu}T_{\mu \nu}.
\end{eqnarray}
Furthermore, using the known relations for Eq. (\ref{eqn:Eeqrw})
\begin{eqnarray}
\label{eqn:Eeqrw2}
R_{0}^{0}
=
\frac{1}{c^2}
\frac{\partial^2 \phi}{\partial x^{i2}},
\end{eqnarray}
\begin{eqnarray}
\frac{8 \pi G}{c^4}(T_{0}^{0}-\frac{1}{2}\delta_{0}^{0}T)
=\frac{8 \pi G}{c^4}\frac{1}{2}(
\mu_{\rm m}
c^2),
\end{eqnarray}
and from Eqs. (\ref{eqn:New1})-(\ref{eqn:New3}), we obtain the Newtonian equation
\begin{eqnarray}
\label{eqn:Neweq}
\frac{\partial^2 \phi}{\partial x^{i2}}=4 \pi G \mu_{\rm m}.
\end{eqnarray}
\par
From Eq. (\ref{eqn:New1}), the term $\delta g^{00} \sqrt{-g}$ in Eq. (\ref{eqn:vari}) is approximated as (higher-order terms with respect to $1/c$ in $\sqrt{-g}$ are neglected)
\begin{eqnarray}
\delta g^{00}\sqrt{-g}  \approx + \delta( \frac{
2 
\phi}{c^2}).
\end{eqnarray}
We then approximate the action functional for directly leading to the Newtonian equation as follows.
Because the action functional for the matter is linear with respect
to $\phi$,
this action is approximated as
\begin{eqnarray}
\label{eqn:SmN}
S_{\rm m}^{(\rm N)}=
- 
\frac{-c^3}{16\pi G}
(\frac{
2 
}{c^2})
\frac{8\pi G}{c^4} \int\frac{1}{2} (
\mu
_{\rm m} c^2) \phi dV_4.
\end{eqnarray}
Meanwhile, we approximate the following action functional of the gravity, which is consistent with the above equation (the factor 1/2 appears considering the variational of both $\partial^2 \phi / \partial x^{i2}$ and $\phi$ with respect to $\phi$),
with the integration by parts
\begin{eqnarray}
\label{eqn:SgN}
S_{\rm g}^{(\rm N)}=
- 
\frac{-c^3}{16\pi G}
(\frac{
2 
}{c^2})
\frac{1}{c^2}
\frac{1}{2}
\int (\frac{\partial \phi}{\partial x^{i}})
     (\frac{\partial \phi}{\partial x^{i}})
dV_4.
\end{eqnarray}
By variational calculus with respect to
$\phi$,
the above action functionals $S_{\rm g}^{(\rm N)}$ and $S_{\rm m}^{(\rm N)}$ provide the Newtonian equation given by Eq. (\ref{eqn:Neweq}).
\par
In the Newtonian approximation within the Einstein scheme, the energy-momentum tensor has the
similar
form as that in
Eq. (\ref{eqn:CE2SE}) for
the static electric field.
Using the notation
$\nabla =(\partial x^1,\partial x^2,\partial x^3)$,
the static energy is written by
\begin{eqnarray}
\label{eqn:EnN}
\nonumber
E^{(\rm N)}=-\frac{1}{8 \pi}\int dr (4 \pi) r^2 (-\nabla \phi) \cdot (-\nabla \phi)
\end{eqnarray}
\begin{eqnarray}
=\frac{1}{2} \int dr (4 \pi) r^2 \mu_{\rm m} \phi.
\end{eqnarray}
\par
As in Subsections \ref{sec:Sec22}, \ref{sec:Sec31} and \ref{sec:Sec32}, the 3D sphere, which is the approximation of the cubic element in 3D space with the cutoff length $\Delta$, has the radius $R_{0}=\Delta/2$ and mass density $\mu_{\rm m}$ in Eq. (\ref{eqn:DmaMQss}) of the fundamental particle. The mass $M$ in Eq. (\ref{eqn:DchMQge}) is the product between $\mu_{\rm m}$ and the volume of the 3D sphere.
From Eq. (\ref{eqn:Neweq}), the gravitational potential, which corresponds to Eq. (\ref{eqn:poEto}) in the electrodynamical case, becomes
\begin{eqnarray}
\label{eqn:poto}
\phi(r)=-G\frac{M}{r}
\hspace{3ex}\mbox{ for } r>R_0.
\end{eqnarray}
The similar correspondence to Eq. (\ref{eqn:poEti2})
for $r \leq R_0$ gives
\begin{eqnarray}
\label{eqn:poti2}
\nonumber
\phi(r)=
-G \frac{4 \pi}{6} r^2 \mu_{\rm m}
+G \frac{4 \pi}{6} R_0^2 \mu_{\rm m}
-G\frac{M}{R_{0}}
\hspace{1.0ex}\mbox{ for } r \leq R_0.
\end{eqnarray}
\begin{eqnarray}
\end{eqnarray}
Subsequently, from
Eqs. (\ref{eqn:EnN}), (\ref{eqn:poti2}) and $M$ in Eq. (\ref{eqn:DchMQge}),
we have the following gravitational self-energy in the case of the continuum theory
\begin{eqnarray}
\label{eqn:CSE}
\nonumber
E^{(\rm N)}=\frac{1}{2}\int_0^{R_{0}} dr (4 \pi) r^2 \mu_{\rm m} \phi (r)
\end{eqnarray}
\begin{eqnarray}
=-\frac{1}{2}
(\frac{4}{5} \frac{GMM}{R_{0}})
=-\frac{1}{2}
(\frac{4}{5} \frac{GMM}{\Delta/2}).
\end{eqnarray}
\par
In contrast to the above analysis, we next study the self-energy in the Einstein field equations, using the step-function-type basis functions.
We also use the above 3D sphere with the radial cutoff length $R_{0}=\Delta/2$ related to the cutoff length $\Delta$. The mass density $\mu_{\rm m}$ in Eq. (\ref{eqn:DmaMQss}) of the fundamental particle provides the mass $M$ in Eq. (\ref{eqn:DchMQge}).
As in Subsec. \ref{sec:Sec22}, we divide three-dimensional space, containing the above sphere (with the radius $R_{0}=\Delta_{\rm h}$) centered at the origin in spherical coordinates, into shells (the number of cells enclosing the central sphere is $N_r-1$). The radial width of each shell is $\Delta_{\rm h}$, which is equal to the radius of the sphere $R_{0}$.
\par
To use the basis functions in Subsec. \ref{sec:Sec22},
the action functional for the gravity in Eq. (\ref{eqn:SgN}) with the factor
\begin{eqnarray}
\gamma_{\rm g}=
\frac{-c^3}{16\pi G} \frac{
2 
}{c^2}\frac{1}{c^2}
\Delta 
x^0,
\end{eqnarray}
($\Delta x^0$
is the time interval and can be dropped for the present static case) is rewritten as
(considering the form 
$\nabla^2 \phi-4\pi G \mu_{\rm m}=0$ of Eq. (\ref{eqn:Neweq}))
\begin{eqnarray}
\label{eqn:AcNgr}
\nonumber
S_{\rm g}^{(\rm N)}=
\frac{\gamma_{\rm g}}{2} \int
dxdydz
\end{eqnarray}
\begin{eqnarray}
\nonumber
\times
[
\phi(
x,y,z
)(
 \frac{\partial ^2}{\partial^2 x}
+\frac{\partial ^2}{\partial^2 y}
+\frac{\partial ^2}{\partial^2 z}
) \phi(
x,y,z
)]
\end{eqnarray}
\begin{eqnarray}
=
\frac{\gamma_{\rm g}}{2} \int dr (4 \pi r^2) \phi^{\rm}(r)(
 \frac{d ^2}{d^2 r}
 +\frac{2}{r}\frac{d}{d r}
 )\phi(r).
\end{eqnarray}
Using the transformed potential $u(r)$ in Eq. (\ref{eqn:tranu}) for spherical coordinates,
we find
\begin{eqnarray}
\label{eqn:Acrr}
S_{\rm g}^{(\rm N)}
=\frac{-\gamma_{\rm g}}{2} \int dr (4 \pi)
\frac{du(r)}{dr}
\frac{du(r)}{dr}.
\end{eqnarray}
The above wave function $u(r)$ is then expressed in terms of the step-function-type basis functions in Eq. (\ref{eqn:BasD}).
From Eqs. (\ref{eqn:BasE}) and (\ref{eqn:Acrr}),
we have
($k, K=$1, 2,…, $N_{{r}+1}$)
\begin{eqnarray}
\label{eqn:acGG}
\nonumber
S_{\rm g}^{(\rm N)}
=\sum_{k,K} 
[\frac{-\gamma_{\rm g}}{2} \int dr (4 \pi) 
u_k u_K
(\frac{\partial \tilde{\Omega}^{E}_k (r)}{d r}
\frac{\partial \tilde{\Omega}^{E}_K (r)}{d r})
].
\end{eqnarray}
\begin{eqnarray}
\end{eqnarray}
Similar to the action in Eq. (\ref{eqn:DEQg}) for the electric field, the above action
becomes
\begin{eqnarray}
\nonumber
\label{eqn:DEg}
S_{\rm g}^{(\rm N)}=\frac{-4\pi}{2} \gamma_{\rm g} \frac{1}{\Delta_{\rm h}}
\sum_{k,K}
(-u_k u_{K-1} \delta_{k,K-1}
\end{eqnarray}
\begin{eqnarray}
+2u_k u_K \delta_{k,K}-u_k u_{K+1} \delta_{k,K+1}).
\end{eqnarray}
\par
Meanwhile,
using $u(r)$ in Eq. (\ref{eqn:tranu}),
the action functional of the matter in Eq. (\ref{eqn:SmN}) for spherical coordinates becomes
(considering the form 
$\nabla^2 \phi-4\pi G \mu_{\rm m}=0$ of Eq. (\ref{eqn:Neweq}))
\begin{eqnarray}
\nonumber
S_{\rm m}^{(\rm N)}=
-4\pi G \gamma_{\rm g} \int dr (4 \pi) r^2
\mu_{\rm m}
\phi(r)
\end{eqnarray}
\begin{eqnarray}
=-4\pi G \gamma_{\rm g} \int dr (4 \pi) r
\mu_{\rm m}
u(r).
\end{eqnarray}
By the expression
of $u(r)$ given by Eq.
(\ref{eqn:BasE}) 
in terms of basis functions denoted in Eq. (\ref{eqn:BasD}), the above action
is written by
\begin{eqnarray}
S_{\rm m}^{(\rm N)}=
-4\pi G \gamma_{\rm g} \int dr (4 \pi) r
\mu_{\rm m}
\sum_k u_k {\tilde\Omega}_k (r)
\end{eqnarray}
\begin{eqnarray}
\label{eqn:DEm}
=-4\pi G \gamma_{\rm g} 
\sum_k u_k
\frac{3}{2}\frac{M}{\Delta_{\rm h}} \delta_{k,1}.
\end{eqnarray}
\par
From
Eqs. (\ref{eqn:DEg}) and (\ref{eqn:DEm}), the variational calculus with respect to $u_k$
\begin{eqnarray}
\delta S_{\rm g}^{(\rm N)} + \delta S_{\rm m}^{(\rm N)}=0,
\end{eqnarray}
results in
\begin{eqnarray}
\label{eqn:SQEGG}
\frac{1}{\Delta_{\rm h}}
(u_{k-1}-2u_k+u_{k+1})=
\frac{3}{2}\frac{GM}{\Delta_{\rm h}} \delta_{k,1}.
\end{eqnarray}
Corresponding to the electric field case in Eqs. (\ref{eqn:DeQOB6}) and (\ref{eqn:DeQOB7}),
the solution
obtained for $k > 1$ is
\begin{eqnarray}
\label{eqn:DeNOB7}
u_{k}=-GM
\hspace{4ex} \mbox{ for } k > 1.
\end{eqnarray}
Equations
(\ref{eqn:SQEGG}) 
and (\ref{eqn:DeNOB7})
provide the solution
for $k=1$:
\begin{eqnarray}
\label{eqn:DeNIB}
u_1
=\frac{-1}{2}(
\frac{3}{2}GM
+GM)
=\frac{-5}{4}GM
\hspace{4ex} \mbox{ for } k=1.
\end{eqnarray}
\par
Consequently, 
from
Eqs. (\ref{eqn:tranu}),
(\ref{eqn:EnN})
and (\ref{eqn:DeNIB}) with
$r_1=\Delta_{\rm h} /2=
(\Delta /2)/2$,
we derive the following classical curvature self-energy in the region with the radial cutoff length
$\Delta_{\rm h}=R_{0}$ (related to the cutoff length $\Delta$)
and the mass
$M=(4 \pi/3) (\Delta_{\rm h})^3(\mu_{\rm m})$
(in Eqs. (\ref{eqn:DmaMQss})-(\ref{eqn:DchMQge}))
\begin{eqnarray}
\label{eqn:EOmega}
\nonumber
E^{(\rm N)}_{\Omega}
=
\frac{1}{2}
\frac{4 \pi}{3}
(\Delta_{\rm h})^3
\mu_{\rm m}
\frac{(-5)}{4}GM
(\frac{1}
{r_1}
)
\end{eqnarray}
\begin{eqnarray}
\nonumber
=\frac{-1}{2}(
\frac{5}{2} 
\frac{GMM}{2r_1})
=\frac{-1}{2}(
\frac{5}{2} 
\frac{GMM}{\Delta_{\rm h}})
=\frac{-1}{2}(
\frac{5}{2} 
\frac{GMM}{\Delta/2}).
\end{eqnarray}
\begin{eqnarray}
\end{eqnarray}

\section{Relationship between the subtracted self-energy and
vacuum
energy constant
$\Lambda$}
\label{sec:Sec4}

The continuum relativistic theory requires that a fundamental particle be considered pointlike, and the radius of the pointlike particle leads to ultraviolet divergences. However, our formalism can obtain finite self-energies by expressing fields in terms of the step-function-type basis functions.
As in Subsec. \ref{sec:Sec21}, the self-energy subtracted from the energy of the rest mass is included in
vacuum
energy expressed in terms of the constant $\Lambda$ (cosmological constant).
The self-energy calculated using the step-function-type basis function with the cutoff length $\Delta$ is not so different from that calculated by the continuum theory with the same cutoff length as was shown in Sec. \ref{sec:Sec3}.
For a fundamental particle, the self-energy caused by the classical electrodynamics was proportional to $1/\Delta$.
This self-energy is reduced to the following logarithmic form by quantum corrections \cite{YuKa,BJDR}
($\hbar$=$h/(2 \pi$) with $h$ being the Planck constant)
\begin{eqnarray}
\label{eqn:SEQED}
E^{(\rm e)}_{\rm Q}
=
\frac{3}{4\pi}(\frac{e^2}{4\pi
\hbar c}) m_{\rm E}c^2 
\{
\ln
[
\frac
{(\hbar c / \Delta)^2 }
{(m_{\rm E}c^2)^2}
]
+\frac{1}{2}
\}.
\end{eqnarray}
($\Delta$ is the cutoff length and $m_{\rm E}$ is the rest mass of an electrodynamically interacting fundamental particle.)
This reduction of the Coulomb-type self-energy also occurs in chromodynamics with the asymptotic freedom at short distances
\cite{Gross,Poli,PRam}.
However, the more strong divergence of the self-energy at a quantum level appears in the Higgs boson case.
(Fundamental particles (quarks) gain the mass through the coupling to the Higgs field.)
The gravitational field is treated at a classical level, because the cutoff length below in this section is longer than the Planck scale and the gravitational strength is small.
Under such a small gravitational field, our treatment in this section relates
vacuum
energy to the Higgs boson self-energy, which is dominant among other interactions at a quantum level.
This relation (between
vacuum
energy and the Higgs self-energy) determines the cutoff length, reproducing today's tiny value of the cosmological constant $\Lambda$ as described below.
\par
It is known that the matter is mainly composed of protons.
The averaged energy of the rest mass
of the fundamental particles
is
$m_{\rm E}c^2 \approx 3.23$ [MeV].
Considering that the contribution from the mass of the fundamental particles to that of a proton is very small, we set
\begin{eqnarray}
\label{eqn:CuBtE}
\nonumber
\gamma_{\rm E}=\frac
{\mbox{energy of proton}}
{\mbox{energy of fundamental particles}}
\end{eqnarray}
\begin{eqnarray}
\nonumber
=\frac
{\mbox{energy density of matter}}
{\mbox{energy density of fundamental particle}}
\approx 96.7.
\end{eqnarray}
\begin{eqnarray}
\end{eqnarray}
To derive the cutoff length, the ratio $\gamma_{\Lambda}$ is defined by
\begin{eqnarray}
\label{eqn:CuCtE}
\nonumber
\gamma_{\Lambda}=\frac{\mbox{energy density of
vacuum energy with $\Lambda$
}}
{\mbox{energy density of matter}}
\end{eqnarray}
\begin{eqnarray}
=\frac{\mbox{self-energy density}}
{\mbox{energy density of matter}}
\approx
\frac{0.73}{0.04}.
\end{eqnarray}
\par
The fundamental particle (quark) mass $m_{\rm E}$ is due to the coupling to the Higgs field with the coupling constant $\lambda_f$ written by
\begin{eqnarray}
\lambda_f=\frac{\sqrt{2}}{v} m_{\rm E},
\end{eqnarray}
where $v$ is the vacuum expectation value of the symmetry-broken Higgs field.
The Higgs self-energy $E_{\rm H}$
(included in
vacuum
energy)
for the cutoff length $\Delta$ and $\lambda_f$ above is written by
\begin{eqnarray}
\label{eqn:EHa}
\nonumber
|E_{\rm H}|=\frac{\sqrt{2}}{4 \pi} \lambda_f \frac{\hbar c}{\Delta}
c^2
\end{eqnarray}
\begin{eqnarray}
=\frac{\sqrt{2}}{4 \pi} \frac{\sqrt{2}}{v} m_{\rm E} c^2 \frac{\hbar c}{\Delta}
=\frac{1}{2 \pi} \frac{1}{v} m_{\rm E} c^2 \frac{\hbar c}{\Delta}.
\end{eqnarray}
Meanwhile, from Eqs. (\ref{eqn:CuBtE}) and (\ref{eqn:CuCtE}), it follows that
\begin{eqnarray}
\label{eqn:EHb}
\frac{|E_{\rm H}|}{m_{\rm E} c^2}=\gamma_{E}\gamma_{\Lambda}.
\end{eqnarray}
Combining Eqs. (\ref{eqn:EHa}) and (\ref{eqn:EHb}), we have
\begin{eqnarray}
\label{eqn:EHc}
\Delta=\frac{1}{\gamma_{E}}\frac{1}{\gamma_{\Lambda}}
\frac{1}{2 \pi} \frac{\hbar c}{v} .
\end{eqnarray}
\par
Because
$v \approx 246$ [GeV],
we derive the cutoff length
$\Delta \approx$
$7.2 \times 10^{-8}$
[fm],
which corresponds to
$\approx$
$2.7 \times 10^6$
[GeV], that is,
$\Delta_{\rm h} \approx$
$3.6 \times 10^{-8}$
[fm] corresponding to
$\approx$
$5.5 \times 10^6$
[GeV].
\par
Even if modifications of values or definitions are required for the above calculations,
we obtain a similar value of $\Delta$.
Owing to the above cutoff length $\Delta$ derived, the field theory
may be
advanced without ultraviolet divergences.
When the renormalization is difficult, the analysis is possible by using the cutoff length. 
In general, $g_{\mu\nu}$ is written by
$g_{\mu\nu}(x)=\eta_{\mu\nu}(x)+h_{\mu\nu}(x)$ as given in Eq (\ref{eqn:New0}),
and the
tensor field $h_{\mu\nu}(x)$
is expressed by
\begin{eqnarray}
h_{\mu\nu}(x)=\sum_p h_{\mu\nu p}{\tilde{\Omega}}^{4}_{p}(x),
\end{eqnarray}
where the coefficients $h_{\mu\nu p}$ are tensor elements,
and ${\tilde{\Omega}}^{4}_{p}(x)$ is the four-dimensional basis function, which
takes the value of unity (one) in a hyper-octahedron with the index $p$ in four-dimensional space-time and vanishes out of the hyper-octahedron.
Namely, the present theory divides space-time continuum of classical general relativity into pieces (hyper-octahedrons). The expression of wave functions in terms of step-function-type basis functions restricts degrees of freedom of the wave functions in a hyper-octahedron (cuts off high-frequency contributions), meaning the quantization of space-time in classical general relativity.
In Sec. \ref{sec:Sec3}, it was shown that the quantities calculated using the step-function-type basis functions are similar to the corresponding quantities calculated by using the continuum theory with the cutoff.
The formalism and calculated quantities in the continuum theory are mapped to the corresponding formalism and quantities using step-function-type basis functions.
Then, from Eqs. (\ref{eqn:EHa})-(\ref{eqn:EHc}), the cut-off length used for the step-function-type basis functions is related to Higgs self-energy, which amounts to the vacuum energy expressed by the constant $\Lambda$ (of such as $\Lambda$-CDM).
\par
In the Einstein field equations, the energy of the rest mass is the source of the curvature, and the renormalization (by such as the electrodynamic interaction) subtracts the self-energy from the rest mass.
The subtracted self-energies can be involved in
vacuum
energy constant
$\Lambda$
in the Einstein field equations, as described in Sec. {\ref{sec:Sec2}}.
(Concerning the curvature self-energy by the gravitational coupling between the mass and the produced field, it is known that the general curvature self-energy is not always within the renormalization scheme.)
The relatively large cut-off length (compared to the Planck length) of the present theoretical formalism has an advantage with naturalness that Higgs self-energy is suppressed, and this cutoff is related to the today's tiny vacuum energy expressed by $\Lambda$, without fine tuning. In contrast, the relatively small cutoff at the Planck scale in other models arises the huge Higgs self-energy, which needs the following fine-tuning. In a highly precise fine-tuning, the huge Higgs self-energy for the cutoff at the Planck scale is cancelled by another physical quantity to adjust the Higgs mass.
Furthermore, the Planck energy composing $\Lambda$ (cosmological constant) in other models need fine tuning to obtain the present tiny value of $\Lambda$ by the cancellation from such as a huge Higgs self-energy for the cutoff at the Planck scale. Therefore, the present formalism provides an answer to solve the fine-tuning problem.
The present formalism also have a possibility to offer a fundamental physical theory, predicting cut-off length that may play a role of a fundamental physical constant, if experimentally observed.
The present model has another merit that the initial universe has no possibility to form a black hole because of the relatively large cut-off length unlike a cutoff at the Planck scale near the black hole size of the whole universe.
Furthermore, the gravity in the present formalism is weak compared to the other fundamental interactions, and space-time coordinates do not largely deviate from classical numbers.
Moreover, the present theoretical
vacuum
energy constant
$\Lambda$
decreases to today's order of magnitude expressing
vacuum
energy density, holding the ratio of
vacuum
energy density to the energy density of the matter.
\par
If the universe expansion is matter dominated (in the present case
vacuum
energy caused by the self-energy has the same property of the matter),
vacuum
energy constant $\Lambda$
seems to be proportional to
$t_{\rm U}^{-2}$
(at least) at present,
where $t_{\rm U}$ is the age of the universe at each point in time.
This is due to the known fact that the solution of the Friedmann equation indicates the universe scale $a(t_{\rm U})$
as a function of $t_{\rm U}$
to be $a(t_{\rm U}) \propto t_{\rm U}^{2/3}$, that is,
$[a(t_{\rm U})]^{-3}$
$\propto$
$t_{\rm U}^{-2}$. 
The matter density $\rho_{\rm m}$ (we can include the dark matter and
vacuum
energy into the matter) is written by $\rho_{\rm m} \propto$
$[a(t_{\rm U})]^{-3}$,
which leads to
$\rho_{\rm m}$
$\propto$
$t_{\rm U}^{-2}$
and $\Lambda \propto$
$t_{\rm U}^{-2}$,
because
vacuum
energy expressed by
vacuum
energy constant
$\Lambda$
has the relation $\Lambda \propto \rho_{\rm m}$ in our scheme.
This is the reason why
vacuum
energy constant
$\Lambda$
seems to be proportional to
$t_{\rm U}^{-2}$.
In contrast, at the Planck scale the wave packet size
for the mass of the Planck energy ($\approx 10^{19}$ [GeV]), which seems to be the whole energy of the universe, is equal to the gravitational-based radius, and the conditions between the above two cases are quite different.
When the present model is generalized to the early universe, the initial size of the universe is the cut-off length of the present model.
Because $\Lambda \propto t_{\rm U}^{-2}$ mentioned above diverges in the limit as $t_{\rm U} \rightarrow 0$ ($t_{\rm U}$ is larger than the cut-off length), the early universe expands at a rapid rate, although the expansion rate is different from the exponential expansion of inflation models.
\par
Finally, we add that owing to the nonperiodic boundary condition,
the zero-point energy for the candidate of
vacuum
energy constant
$\Lambda$
is not seen in the present formalism.
Let us consider a simple action for the wave function $\Phi(x)$
\begin{eqnarray}
S_{\rm B}=\frac{1}{2} \int dx (\frac{d \Phi(x)}{dx} \frac{d {\Phi(x)}}{dx}).
\end{eqnarray}
The wave function in terms of the basis functions in
Eq. (\ref{eqn:BasD}) with the lattice spacing
$\Delta$
is given by
\begin{eqnarray}
\Phi(x)=\sum_k \Phi_k {\tilde\Omega}_{k}^{E}(x).
\end{eqnarray}
Similar to Eq. (\ref{eqn:DEQg}), we
write the action functional
\begin{eqnarray}
\label{eqn:ElAc}
\nonumber
S_{\rm B}=\frac{1}{2}\sum_{k,K}
\int dx
(\Phi_k \Phi_K
\frac{d {\tilde\Omega}_{k}^{E}(x)}{dx}
\frac{d {\tilde\Omega}_{K}^{E}(x)}{dx})
\end{eqnarray}
\begin{eqnarray}
=\frac{1}{2}\frac{1}{\Delta}\sum_{k,K}
(\Phi_k \Phi_{K})(-\delta_{k-1,K}+2\delta_{k,K}-\delta_{k+1,K}).
\end{eqnarray}
The variational calculus with respect to $\Phi_k$
\begin{eqnarray}
\delta S_{\rm B}=0,
\end{eqnarray}
yields
\begin{eqnarray}
\frac{1}{\Delta}(\Phi_{k-1}-2\Phi_{k}+\Phi_{k+1})=0,
\end{eqnarray}
which provides
\begin{eqnarray}
\frac{1}{{\Delta}^2}(\Phi_{k-1}-2\Phi_{k}+\Phi_{k+1})=0,
\end{eqnarray}
where $k, K=1,2,...,N_{x}$. 
For the above equation, the following boundary conditions on the wave function are imposed 
($N_{x}+2$ is the number of lattice points, and lattice indices of the boundary points are denoted by $k=0, N_{x}+1$):
\begin{eqnarray}
\label{eqn:BCon}
\Phi_0=0,
\hspace{4ex}
\Phi_{N_{x}+1}=0.
\end{eqnarray}
\par
Similar to the classical vibrational case \cite{SlaFra}, the eigenvector for a
diagonalization of the action is expressed as
\begin{eqnarray}
\Phi_{K}=\frac{1}{C_{\rm N}}\sin(\frac{kK \pi}{N_{x}+1}),
\end{eqnarray}
where $C_{\rm N}$ is
a normalization constant.
Then, the element $S_{{\rm B}, k}$ of the action in Eq. (\ref{eqn:ElAc}) is diagonalized giving
\begin{eqnarray}
\nonumber
S_{{\rm B},k}=
\frac{1}{2}\frac{1}{\Delta}
\sum_{K}(-\delta_{k-1,K}+2\delta_{k,K}-\delta_{k+1,K})\Phi_{K}
\end{eqnarray}
\begin{eqnarray}
\nonumber
=
\frac{1}{2\Delta}\frac{1}{C_{\rm N}}
\end{eqnarray}
\begin{eqnarray}
\nonumber
\times\{ -\sin[\frac{k(k-1) \pi}{N_{x}+1}]
+2\sin[\frac{kk \pi}{N_{x}+1}]
-\sin[\frac{k(k+1) \pi}{N_{x}+1}] \}
\end{eqnarray}
\begin{eqnarray}
\nonumber
=\frac{1}{\Delta}\frac{1}{C_{\rm N}}
[1-\cos(\frac{k \pi}{N_{x}+1})]
\sin(\frac{kk \pi}{N_{x}+1})
=\eta_{k}\delta_{k,K}\Phi_{K},
\end{eqnarray}
\begin{eqnarray}
\end{eqnarray}
yielding the eigenenergies
\begin{eqnarray}
\eta_{k} \propto 1-\cos(\frac{k \pi}{N_{x}+1}),
\end{eqnarray}
with $
k
=1, 2,\cdot\cdot\cdot,N_{x}$.
Consequently, 
the zero-point energy for the candidate of
vacuum
energy is not seen in the present system because of the boundary condition in Eq. (\ref{eqn:BCon}).
(Similarly, eigenvalues in higher dimensions are obtained \cite{Fuku14,Fuku17}.)
\par
We note that when the system is considered using a box normalization, in which wave functions are defined in a box with periodic boundary conditions at box surfaces, eigenvalues may have zero-point energies. However, physical quantities such as the transition amplitude are calculated without using the zero-point energies by expressing plane waves in the form of complex exponential functions.
The zero-point energies dropped in this case may not be included in
vacuum
energy, because the zero-point energies appeared here are due to the non-vanishing periodicity (which seems to be lack in the real expanding universe) in
approximate
calculation manipulations.
\par
As the dark matter, we considered the classical solution with quantum field fluctuations in chromodynamics in our previous paper \cite{Fuku17}.
Although the Big Bang is out of the scope of this paper, an expansion may arise such as the Big Bang like the vaporization of water in vacuum by absorbing heat.
\par

\section{Conclusions}
\label{sec:Sec5}

For the renormalization of the mass, we have considered the subtracted
self-energies, which act as sources of the curvature in the Einstein field equations. It was shown that this consistency is satisfied by including these self-energies into
vacuum
energy expressed by the constant $\Lambda$.
The self-energies in electrodynamics and Einstein field equations were investigated by using the ultraviolet cutoff length. The field theory, which was developed by the present authors, expresses wave functions in terms of the step-function-type basis functions to cut off oscillations at short distances.
In the other continuum theory, we used the same cutoff length as that used for the former theory.
From the examination, the continuum theory with the cutoff length is effective. Classical self-energies are reduced to logarithmic forms by quantum corrections, and the quadratic Higgs self-energy is dominant at a quantum level.
The cutoff length was determined to reproduce the observed
vacuum
energy constant
$\Lambda$,
using the self-energy derived from the above cutoff theories.
The derived
vacuum
energy expressed by the constant
$\Lambda$
is of the order of the matter (composed of the conventional matter such as atoms and dark matter), showing that
vacuum
energy constant
$\Lambda$
has today's tiny value.

\end{multicols}


\begin{thebibliography}{}

\bibitem{Lamb}
W. E. Lamb, Jr. and R. C. Retherfold, Phys. Rev. {\bf 72}, 339
(1947).
\bibitem{Weiss}
V.~S.~Weisskopf, Kongelige Danske Videnskabernes Selskab, Mathematisk-fysiske Meddeleser {\bf XIV}, No. 6 (1936).
\bibitem{Weiss39}
V.~S.~Weisskopf, Phys. Rev. {\bf 56}, 72 (1939).
\bibitem{Tomo46}
S.~Tomonaga, Prog. Theo. Phys. {\bf 1}, 27 (1946).
\bibitem{Tomo48}
S.~Tomonaga, Phys. Rev. {\bf 76}, 224 (1948).
\bibitem{Schwin}
J.~Schwinger, Phys. Rev. {\bf 76}, 790 (1949).
\bibitem{Feyn}
R.~P.~Feynman,
Phys. Rev.
{\bf 76}, 769 (1949).
\bibitem{Dys}
F.~J.~Dyson, Phys. Rev. {\bf 75}, 1736 (1949).
\bibitem{Fuku84}
K.~Fukushima, Phys. Rev. D {\bf30}, 1251 (1984).
\bibitem{Fuku14}
K.~Fukushima and H.~Sato, Bulg. J. Phys. {\bf 41}, 142 (2014); \\
arXiv:1402.0450 (arXiv:1402.0450v5). \\
Freely available at \\
http://www.bjp-bg.com/papers/bjp2014\_2\_142-171.pdf
\bibitem{Fuku16}
K.~Fukushima and H.~Sato, Bulg. J. Phys. {\bf 43}, 30 (2016); \\
arXiv:1402.0450 (arXiv:1501.04837v6). \\
Freely available at \\
http://www.bjp-bg.com/papers/bjp2016\_1\_030-044.pdf
\bibitem{Fuku17}
K.~Fukushima and H.~Sato,
Int. J. Mod. Phys. A {\bf 32}, 1730017 (2017); arXiv:1705.03767 (arXiv:1705.03767v6)
\bibitem{FemMW}
A.~R.~Mitchell and R.~Wait, {\it The Finite Element Method in Partial
Differential Equations}, (John Wiley \& Sons, New York, 1977).
\bibitem{BenMS}
C.~M.~Bender, K.~A.~Milton and D.~H.~Sharp, Phys.\ Rev.\ Lett. {\bf 51}, 1815 (1983).
\bibitem{LanL}
L.~D.~Landau and E.~M.~Lifshitz, {\it The Classical Theory of Fields}, 4th revised English edition, (Elsevier, Amsterdam, 1951).
\bibitem{Eins}
A.~Einstein, Sitzungsberichte der Koniglich Preussischen Akademie der Wissenschaften Berlin. part 1, 142 (1917).
\bibitem{Fried}
A.~Friedmann, Z. Phys. {\bf 10}, 377 (1922). 
\bibitem{Carr92}
S.~M.~Carroll, H.~P.~William and E.~L.~Turner, The cosmological constant Annual Review of Astronomy and Astrophysics  {\bf 30}, 499 (1992). 
\bibitem{Carr01}
S.~M.~Carroll, The Cosmological Constant Living Reviews in Relativity {\bf 4}, 1 (2001). 
\bibitem{Carr04}
S.~Carroll, {\it Spacetime and Geometry}, (Addison Wesley, San Francisco, CA, 2004) p. 171.
\bibitem{Riess}
A.~Riess {\it et al.}, The Astronomical Journal {\bf 116}, 1009 (1998).
\bibitem{Baker}
J.~C.~Baker {\it et al.}, Monthly Notices of the Royal Astronomical Society {\bf 308}, 1173 (1999).
\bibitem{Perl}
S.~Perlmutter {\it et al.}, The Astrophysical Journal {\bf 517}, 565 (1999).
\bibitem{Peeb}
P.~J.~E.~Peebles and B.~Ratra, Rev. Mod. Phys. {\bf 75}, 559 (2003).
\bibitem{Padman}
T.~Padmanabhan, Phys. Rep. {\bf 380}, 235 (2003). 
\bibitem{Tegm}
M.~Tegmark et al. Phys. Rev. D {\bf 69}, 103501 (2004).
\bibitem{Lomb}
L.~Lombriser, L.~Lucas and A.~Nelson, Physics Letters B. {\bf 765}, 382 (2017).
\bibitem{Barr}
J.~D.~Barrow and D.~J.~Shaw, General Relativity and Gravitation {\bf 43}, 2555 (2011).
\bibitem{Rugh}
S.~Rugh and H.~Zinkernagel, Studies in History and Philosophy of Modern Physics {\bf 33}, 663 (2001).
\bibitem{Hobs}
M.~P.~Hobson, G.~P.~Efstathiou and A.~N.~Lasenby, {\it General Relativity: An Introduction for Physicists}, (Cambridge University Press, Cambridge, 2006).
\bibitem{Wein}
S.~Weinberg, Rev. Mod. Phys. {\bf 61}, 1 (1989).
\bibitem{Leut}
H.~Leutwyler, J.~R.~Klauder and L.~Streit, Nuovo Cim. {\bf A66}, 536 (1970). 
\bibitem{BJDR}
J.~D.~Bjorken and S.~D.~Drell, {\it Relativistic Quantum Mechanics}, (McGraw-Hill, New York, 1964).
\bibitem{Rohr62}
F. Rohrlich, Physics Today {\bf 15}, 19 (1962).
\bibitem{Dira38}
P. A. M. Dirac, Proc. Roy. Soc. (London) {\bf A167}, 148 (1938).
\bibitem{Rohr60}
F. Rohrlich, Am. J. Phys. {\bf 28}, 639 (1960)
\bibitem{Rohr61}
F. Rohrlich, Nuovo Cim. {\bf 21}, 811 (1961).
\bibitem{YuKa}
H.~Yukawa and Y.~Katayama, {\it Soryushi-ron (Theory of Elementary Particles)} (in Japanese), (Iwanami, Tokyo, Japan, 1974).
\bibitem{Gross}
D.~J.~Gross and F.~Wilczek, Phys. Rev. Lett. {\bf 30}, 1343 (1973).
\bibitem{Poli}
H.~D.~Politzer, Phys. Rev. Lett. {\bf 30}, 1346 (1973).
\bibitem{PRam}
P.~Ramond, {\it Field Theory: A Modern Primer}, 3rd prn., (Benjamin, MA, 1982).
\bibitem{SlaFra}
J.~C.~Slater and N.~H.~Frank, {\it Mechanics}, (McGraw-Hill, New York, 1947).

\end{thebibliography}
\end{document}